\documentclass{elsart}
\usepackage{graphicx,amssymb}
\usepackage[latin1]{inputenc}
\usepackage[T1]{fontenc}
\usepackage[centertags]{amsmath}
\usepackage{newcent}
\usepackage{newlfont}
\usepackage{subfigure}
\usepackage{psfrag}
\usepackage{amsmath}
\usepackage{amsfonts}
\usepackage{amssymb}
\usepackage{ae}
\usepackage{feynmf}
\usepackage{cancel}
\usepackage{wasysym}
\usepackage{fancyhdr}
\usepackage{mathrsfs}
\usepackage{url}
\usepackage[noadjust,nocompress]{cite}
\usepackage{pstricks}

\newcommand{\di}{\mathrm{d}}

\newcommand{\modulo}[1]{\left |#1\right |}
\newcommand{\sqmodulo}[1]{{\modulo{#1}}^{2}}

\newcommand{\M}{{\mathscr{M}}}

\newcommand{\Obig}[1]{{\mathcal{O}\left (#1\right )}}
\newcommand{\vAP}{{\mathcal{P}}}

\newcommand{\Iplus}{{\mathcal{I}_{+}}}

\begin{document}
\rightline{FNT/T 2008/01}
\rightline{SHEP-08-05}
\begin{frontmatter}
\title{Photon pair production at flavour factories with per mille accuracy}
\author[Pavia,PaviaI]{G. Balossini}
\author[Pavia]{C. Bignamini}
\author[UK]{C.M. Carloni Calame} 
\author[Pavia,PaviaI]{G. Montagna}
\author[PaviaI]{O. Nicrosini}
\author[PaviaI]{F. Piccinini}
\address[Pavia]{Dipartimento di Fisica Nucleare e Teorica, Universit\`a di Pavia, Via A. Bassi 6, 27100 Pavia, Italy.}
\address[PaviaI]{Istituto Nazionale di Fisica Nucleare, Sezione di Pavia, Via A. Bassi 6, \\ 27100 Pavia, Italy.}
\address[UK]{Istituto Nazionale di Fisica Nucleare and School of Physics \& Astronomy, University of Southampton, Southampton S017 1BJ, England.}

\begin{abstract}
We present a high-precision QED calculation, with $0.1\%$ theoretical accuracy, of two photon production in $e^+ e^-$ annihilation, as required by more and more accurate 
luminosity monitoring at flavour factories. The accuracy of the approach, which is based 
on the matching of exact next-to-leading order corrections with a QED Parton Shower 
algorithm, is demonstrated through a detailed analysis of the impact of the various sources of radiative corrections to the experimentally relevant observables. The calculation is implemented in the latest version of the event generator BabaYaga, available for precision simulations of photon pair production at $e^+ e^-$ colliders of moderately high energies.
\end{abstract}
\end{frontmatter}
\section{Introduction}
The precision measurement of the hadron production cross section in $e^+ e^-$ annihilation at flavour factories, such as $\Phi$, $\tau$-charm and $B$-factories, requires a detailed knowledge of the collider luminosity \cite{proc}. It can be derived by counting the number of events of a given reference process and normalizing this number to the corresponding theoretical cross section \cite{1}. It follows that, in order to maintain small the total luminosity error given by the sum in quadrature of the relative experimental and theoretical uncertainty, the reference process must be a reaction with high statistics and calculable with an accuracy as high as possible. For this reason, the luminosity monitoring processes employed at flavour factories are QED processes, namely Bhabha scattering, 
two photon and muon pair production. In particular, at DA$\Phi$NE, VEPP-2M and PEP-II
 the large-angle Bhabha process is primarily used and the other reactions are measured as cross checks \cite{1,2,3}, while at CESR \cite{3a} all the three processes are considered and the luminosity is derived as an appropriate average of the measurements of the three QED reactions.

At DA$\Phi$NE, a comparison between the luminosity measurement using Bhabha events and
the process $e^+ e^- \rightarrow \gamma\gamma$ shows very good agreement, the average
difference in a run-by-run comparison being 0.3\% \cite{3b}. This precision necessarly demands progress on the theory side, since the Monte Carlo (MC) programs used for the simulation of photon pair production, 
{\it i.e.} BabaYaga v3.5~\cite{B,Ca} and BKQED~\cite{BK}, have a theoretical precision of about 
$1\%$. Actually, the original formulation of 
BabaYaga is based on a QED Parton Shower (PS) approach for the treatment of leading logarithmic  (LL) QED corrections and, as such, it lacks the effect of $\Obig{\alpha}$ non-log contributions, which are important to achieve a precision at the per mille level. On the other hand, the generator BKQED relies on an exact $\Obig{\alpha}$ diagrammatic calculation, therefore neglecting the contribution of higher-order LL corrections, which have been already demonstrated to be necessary for $\Obig{0.1\%}$ luminosity monitoring at flavour factories \cite{B,B1,2}. Because of this motivation, the aim of the present paper is to describe a high-precision calculation of photon pair production in QED, based on the 
matching of exact next-to-leading-order (NLO) corrections with the QED PS algorithm, along the lines of the approach already developed for the Bhabha process in Ref.~\cite{B1}. This will allow a reduction of the theoretical error in luminosity measurements at flavour factories, as demanded, in addition to precision measurements of the hadronic cross section, by improved experimental determinations of the $e^+ e^- \rightarrow \tau^+ \tau^-$ cross section at low energies~\cite{tau}, important for
precision calculations of the anomalous magnetic moment of the muon.  Furthermore, a precise 
knowledge of $e^+ e^-$ annihilation in two photons is of interest for estimates of the background to neutral meson production.
We do not include in our calculation pure weak corrections, which have been computed 
in Ref.~\cite{bs} and turn out to be important at very high energies, well above the energy range
explored by flavour factories. For completeness, it is worth mentioning that an independent calculation, 
including exact $\Obig{\alpha}$ contributions supplemented with higher-order LL terms 
through collinear QED Structure Functions~\cite{sf},  of 
the relevant corrections to $e^+ e^- \rightarrow \gamma\gamma$ at moderately high energies 
was performed in 
Ref.~\cite{aetal}, recently revisited in Ref.~\cite{newgg}.

The outline of the paper is as follows. In Section 2 we 
describe the matching algorithm for 
the $e^+ e^- \rightarrow \gamma\gamma$ process, while in Section 3 we provide numerical results, both for integrated cross sections and differential distributions of experimental interest, in order to discuss the effects of the various sources of radiative corrections and 
provide evidence for the per mille accuracy of the approach. Conclusions and possible 
perspectives are drawn in Section 4.

\section{Theoretical formulation}
The cross section of the photon pair production process, with the additional emission of an arbitrary number of photons, can be written in the LL  approximation as follows
\begin{equation}\label{eq:1}
\di\sigma^{\infty}_{\textrm{LL}}=\Pi^{2}\left(Q^{2},\epsilon\right)\sum_{n=0}^{\infty}\dfrac{1}{n!}
\modulo{\M_{n,\textrm{LL}}}^{2}\di \Phi_{n}\ , 
\end{equation}
where $\Pi\left(Q^{2},\epsilon\right)$ is the Sudakov form factor accounting for the soft-photon (up to an energy equal to $\epsilon$ in units of the incoming fermion energy $E$) and virtual emission, $\epsilon$ is an infrared separator dividing soft and hard radiation and $Q^{2}$ is related to the energy scale of the hard-scattering process. In our calculation, $Q^2$ is fixed to be equal to the squared centre of mass 
(c.m.) energy $s$, by comparing with the exact $\Obig{\alpha}$ calculation of Ref.~\cite{BK}.
$\sqmodulo{\M_{n,\textrm{LL}}}$ is the squared amplitude in LL approximation describing the process with the emission of $n$ additional hard photons, with energy larger than $\epsilon$ in units of $E$, 
with respect to the lowest-order approximation 
$e^+ e^- \to \gamma\gamma$. $\di\Phi_{n}$ is the exact phase space element of the process (divided by the incoming flux factor), with the emission of $n$ additional photons with respect to the Born-like final state configuration. The Sudakov form factor, which is defined as
\begin{equation}
\Pi\left (Q^{2},\epsilon\right )=\exp\left(-\dfrac{\alpha}{2\pi}\mathcal{I}_{+}L\right)\ ,
\end{equation}
where
\begin{equation}\label{eq:1b}
L=\log\dfrac{Q^{2}}{m^{2}}\qquad \Iplus=\int_{0}^{1-\epsilon}\di z\ \vAP\left (z\right)\ ,
\end{equation}
appears in Eq. (\ref{eq:1}) to the second power to account for the 
presence of two charged particles in the initial state. In Eq. (\ref{eq:1b}) $\vAP(z)$ is the
electron $\to$ electron + photon splitting function $\vAP (z) = (1+z^2)/(1-z)$.

The cross section as calculated in Eq. (\ref{eq:1}) has the advantage that the photonic corrections, in LL approximation, are resummed up to all orders in perturbation theory. On the other 
hand, the weak point of Eq. (\ref{eq:1}) is that its  $\Obig{\alpha}$  expansion does not coincide with the exact $\Obig{\alpha}$ (NLO) result. Actually, we have
\begin{eqnarray}
\label{eq:2}
\hskip 62pt \di\sigma^{\alpha}_{\textrm{LL}}&&=\left(1-\frac{\alpha}{\pi}\mathcal{I}_{+}\ln\frac{Q^{2}}{m^{2}}\right)\modulo{\M_{0}}^{2}\di \Phi_{0}+\modulo{\M_{1,\textrm{LL}}}^{2}\di \Phi_{1} \nonumber\\
&&\equiv\left(1+C_{\alpha,\textrm{LL}}\right)\modulo{\M_{0}}^{2}\di \Phi_{0}+\modulo{\M_{1,\textrm{LL}}}^{2}\di \Phi_{1}\ , 
\end{eqnarray}
whereas an exact NLO can be always cast in the form 
\begin{equation}\label{eq:3}
\di\sigma^{\alpha}=\left(1+C_{\alpha,\textrm{SV}}\right)\modulo{\M_{0}}^{2}\di \Phi_{0}+\modulo{\M_{1}}^{2}\di \Phi_{1}\ , 
\end{equation}
where the coefficient $C_{\alpha,\textrm{SV}}$ is equal to the exact squared amplitude of the annihilation process, in the presence of soft and virtual radiative corrections~\cite{BK,aetal}, in units of the 
exact Born squared amplitude $\modulo{\M_{0}}^{2}$, and $\sqmodulo{\M_{1}}$ is the exact squared matrix element of the 
radiative process $e^+ e^- \rightarrow \gamma\gamma\gamma$~\cite{ks}. The matching of the LL and NLO calculation can be obtained considering the correction factors (free of infrared and collinear singularities)
\begin{equation}\label{eq:4}
F_{\textrm{SV}}=1+\left(C_{\alpha,\textrm{SV}}-C_{\alpha,\textrm{LL}}\right)\qquad F_{\textrm{H}}=1+\dfrac{{\modulo{\M_{1}}}^{2}-{\modulo{\M_{1,\textrm{LL}}}}^{2}}{{\modulo{\M_{1,\textrm{LL}}}}^{2}}\ . 
\end{equation}
As can be seen, the exact $\Obig{\alpha}$ cross section as in Eq. (\ref{eq:3}) can be expressed, up to terms of $\Obig{\alpha^{2}}$ and in terms of its LL approximation, as
\begin{equation}\label{eq:5}
\di\sigma^{\alpha}=F_{\textrm{SV}}\left(1+C_{\alpha,\textrm{LL}}\right)\modulo{\M_{0}}^{2}\di \Phi_{0}+F_{\textrm{H}}\modulo{\M_{1,\text{LL}}}^{2}\di \Phi_{1}\ . 
\end{equation}
A similar procedure, repeated to all orders in $\alpha$, leads to the correction of 
Eq.~(\ref{eq:1}), which becomes 
\begin{equation}\label{eq:6}
\di\sigma^{\infty}_{\textrm{matched}}=F_{\textrm{SV}}\Pi^{2}\left(Q^{2},\epsilon\right)\sum_{n=0}^{\infty}\dfrac{1}{n!}\left(\prod_{i=0}^{n}F_{\textrm{H},i}\right)\modulo{\M_{n,\textrm{LL}}}^{2}\di \Phi_{n}\ , 
\end{equation}
where
\begin{equation}\label{eq:7}
F_{\textrm{H},i}=1+\dfrac{{\modulo{\M_{i}}}^{2}-{\modulo{\M_{i,\textrm{LL}}}}^{2}}{{\modulo{\M_{i,\textrm{LL}}}}^{2}}\ , 
\end{equation}
with ${\modulo{\M_{i}}}^{2}$ and ${\modulo{\M_{i,\textrm{LL}}}}^{2}$ squared matrix elements,  exact and in the LL approximation, respectively, relative to the emission of the $i$-th hard 
bremsstrahlung photon. 
The expansion at $\Obig{\alpha}$ of Eq. (\ref{eq:6}) coincides now with the exact NLO cross section 
of Eq.~(\ref{eq:3}) and higher-order LL contributions are the same as in Eq.~(\ref{eq:1}). 

\section{Numerical results}

\subsection{Integrated cross sections: technical tests and radiative corrections}

\begin{figure}
\begin{center}
\includegraphics[width=285pt]{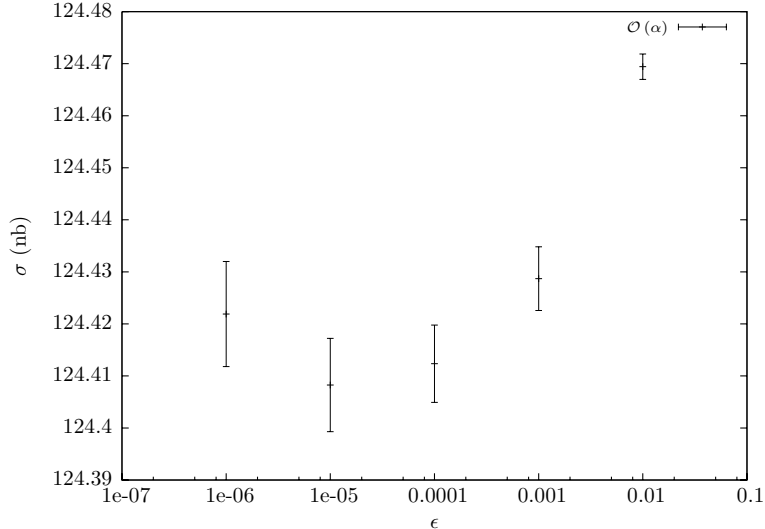}
\end{center}
\caption{$\Obig{\alpha}$ QED corrected cross section of Eq.~(5) 
as a function of the infrared regulator $\epsilon$. The error bars correspond to $1\sigma$ Monte Carlo statistics.}
\label{fig:1}
\end{figure}
\begin{figure}
\begin{center}
\includegraphics[width=285pt]{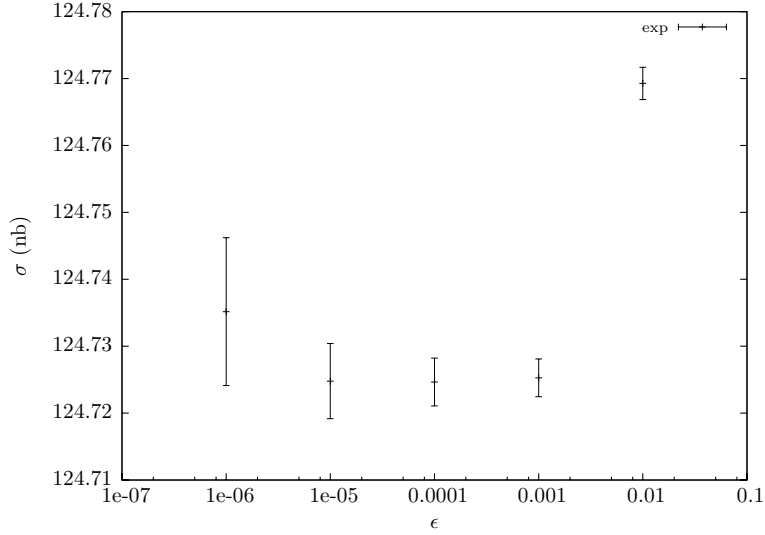}
\end{center}
\caption{The same as Fig. 1 for the matched QED corrected cross section of Eq.~(8).}
\label{fig:2}
\end{figure}

The calculation of QED corrections requires the introduction of the unphysical soft-hard separator 
$\epsilon$. Therefore, the indipendence of the predictions for the QED corrected cross section from variation of such a parameter has to be proved, for sufficiently small $\epsilon$ values. This is succesfully demonstrated, at a precision level of $\sim 0.01$\%,  in Fig.~\ref{fig:1} and Fig.~\ref{fig:2}, which show the cross section of the photon pair production process, obtained according to the exact $\Obig{\alpha}$ cross section of Eq.~(\ref{eq:3}) (Fig.~\ref{fig:1}) and to the matched
cross section of Eq.~(\ref{eq:6}) (Fig.~\ref{fig:2}), as a function of $\epsilon$ from $10^{-2}$ to $10^{-6}$. The numerical results shown in Fig. \ref{fig:1} and Fig.~\ref{fig:2} correspond to 
the following experimental set up, which models, up to a good accuracy, 
the selection criteria adopted by KLOE Collaboration at DA$\Phi$NE~\cite{ad}
\begin{equation}\label{eq:KLOE}
    \left \{
\begin{aligned}
    \sqrt{s} &= 1.02\ \textrm{GeV}\\
    E_{\gamma}^{\textrm{min}} &= 0.3\ \textrm{GeV}   \\
    \vartheta_{\gamma}^{\textrm{min}} &= 45^{\circ}    \\
    \vartheta_{\gamma}^{\textrm{max}} &= 135^{\circ}  \\
    \xi_{\textrm{max}}& =10^{\circ}
\end{aligned}
  \right.
\end{equation}
where $E_{\gamma}^{\textrm{min}}$ is the minimum energy threshold for the detection of at 
least two photons, $\vartheta_{\gamma}^{\textrm{min,max}}$ are the angular acceptance cuts and 
$\xi_{\textrm{max}}$ is the maximum acollinearity between the most energetic and next-to-most 
energetic photon.

As a further test of the approach, we checked that our results for the NLO corrections agree 
at the 0.1\% level with those quoted in Ref. \cite{BK} for the exact $\Obig{\alpha}$ relative 
corrections to the totally inclusive $e^+ e^- \rightarrow \gamma\gamma$ cross section, as a 
function of different c.m. energies.

\begin{table}[!htbp]
\begin{center}
\begin{tabular}{|l|c|c|c|}
\hline
$\sqrt{s}\ \left(\textrm{GeV}\right)$ & $1$ & $3$ & $10$ \\
\hline
\hline
$\sigma$ & $137.53$ & $15.281$ & $1.3753$ \\
\hline
$\sigma_{\alpha}^{\textrm{PS}}$ & $128.55$ & $14.111$ & $1.2529$ \\
\hline
$\sigma_{\alpha}^{\textrm{NLO}}$ & $129.45$ & $14.211$ & $1.2620$ \\
\hline
$\sigma_{\textrm{exp}}^{\textrm{PS}}$ & $128.92$ & $14.169$ & $1.2597$ \\
\hline
$\sigma_{\textrm{exp}}$ & $129.77$ & $14.263$ & $1.2685$ \\
\hline
\hline
$\delta_{\alpha}$ & $-5.87$ & $-7.00$ & $-8.24$ \\
\hline
$\delta_{\infty}$ & $-5.65$ & $-6.66$ & $-7.77$ \\
\hline
$\delta_{\textrm{exp}}$ & $0.24$ & $0.37$ & $0.51$ \\
\hline
$\delta_{\alpha}^{\textrm{NLL}}$ & $0.70$ & $0.71$ & $0.73$ \\
\hline
$\delta_{\infty}^{\textrm{NLL}}$ & $0.66$ & $0.66$ & $0.69$ \\
\hline
\end{tabular}
\end{center}
\caption{Photon pair production cross sections (in nb) to different 
accuracy levels and relative corrections (in per cent) for the set up specified in the text.} 
\label{tab:1}
\end{table}

To quantify the overall impact of QED radiation and, in particular, to evaluate the size of QED contributions at different perturbative and precision levels, we show in Tab. \ref{tab:1} the Born cross section $\sigma$, the $\Obig{\alpha}$ PS and exact cross section, $\sigma_{\alpha}^{\textrm{PS}}$ and $\sigma_{\alpha}^{\textrm{NLO}}$, respectively, 
as well as the all-order PS cross section of Eq. (\ref{eq:1}) $\sigma_{\textrm{exp}}^{\textrm{PS}}$\footnote{The results denoted as $\sigma_{\textrm{exp}}^{\textrm{PS}}$ agree with the 
predictions of BabaYaga v3.5 within 0.1\% accuracy, as we checked explicitly.}  and the matched cross section of Eq. (\ref{eq:6}) $\sigma_{\textrm{exp}}$. The results have been obtained with experimental selection criteria similar to those of Eqs. (\ref{eq:KLOE}), but with c.m. energies $\sqrt{s} = 1,3,10\ \textrm{GeV}$ and with the condition $E_{\gamma}^{\textrm{min}} = 0.3\sqrt{s}$. From these cross section values, it is possible to calculate the relative effect of various QED contributions, namely the contribution of exact $\Obig{\alpha}$ radiative corrections, of higher-order corrections in the $\Obig{\alpha}$ matched PS scheme, of exponentiation with respect to the exact $\Obig{\alpha}$ cross section and, finally, of non-logarithmic terms entering the $\Obig{\alpha}$ cross section and present in the improved PS algorithm. 
The above per cent corrections are shown in Tab. \ref{tab:1} and they can be derived from the cross section values according to the following formulae
\begin{equation*}
\delta_{\alpha}=100\times\dfrac{\sigma_{\alpha}^{\textrm{NLO}}-\sigma}{\sigma}\qquad 
\delta_{\infty}=100\times\dfrac{\sigma_{\textrm{exp}}-\sigma}{\sigma}
\end{equation*}
\begin{equation}\label{eq:deltas}
\delta_{\textrm{exp}}=100\times\dfrac{\sigma_{\textrm{exp}}-\sigma_{\alpha}^{\textrm{NLO}}}{\sigma_{\alpha}^{\textrm{NLO}}}\qquad
\delta_{\alpha}^{\textrm{NLL}}=100\times\dfrac{\sigma_{\alpha}^{\textrm{NLO}}-\sigma_{\alpha}^{\textrm{PS}}}{\sigma_{\alpha}^{\textrm{PS}}}
\end{equation}
\begin{equation*}
\delta_{\infty}^{\textrm{NLL}}=100\times\dfrac{\sigma_{\textrm{exp}}-\sigma_{\textrm{exp}}^{\textrm{PS}}}{\sigma_{\textrm{exp}}^{\textrm{PS}}}\ . 
\end{equation*}

The numerical errors coming from the MC integration are not shown in Tab.~\ref{tab:1} because
they are beyond the quoted digits. From Tab. \ref{tab:1} it can be seen that the exact $\Obig{\alpha}$ corrections, measured by the relative contribution $\delta_{\alpha}$, lower the Born cross section of about $5.9\%$ ($\Phi$ resonance), $7.0\%$ ($J/\psi$ resonance) and $8.2\%$ ($\Upsilon$ resonance). All-order corrections, due to the presence of an arbitrary number of photons and measured by the relative contribution $\delta_{\infty}$, amount to about $5.7\%$ ($\Phi$ resonance), $6.7\%$ ($J/\psi$ resonance) and $7.8\%$ ($\Upsilon$ resonance), showing that the introduction of higher photon multiplicity gives an increasing of the 
$\Obig{\alpha}$ corrected cross section. Such an effect, due to $\Obig{\alpha^{n}L^{n}}$ 
(with $n \geq 2$) terms, is quantified by the contribution $\delta_{\textrm{exp}}$, which is 
a positive correction of about $0.2\%$ ($\Phi$ resonance), $0.4\%$ ($J/\psi$ resonance) and $0.5\%$ ($\Upsilon$ resonance), and, therefore, important in the light of the aimed per mille accuracy. 
On the other hand, also next-to-leading $\Obig{\alpha}$ corrections, quantified
by the  contribution $\delta_{\alpha}^{\textrm{NLL}}$, are necessary at the precision level 
of 0.1\%, since their contribution is of about $0.7\%$, almost indipendently of the c.m. energy. Their
effect is unaltered at the level of 0.1\% by the matching procedure with PS, as can be inferred by 
comparing $\delta_{\alpha}^{\textrm{NLL}}$ with $\delta_{\infty}^{\textrm{NLL}}$. To further corroborate
the precision reached in the cross section calculation, we also evaluated the effect due to the 
most important sub-leading $\Obig{\alpha^2}$ photonic corrections and given by $\alpha^2 L$ 
contributions enhanced by infrared logarithms. Actually, the bulk of such corrections is effectively 
incorporated in our approach, by means of factorization of $\Obig{\alpha}$ next-to-leading terms
with the leading $\Obig{\alpha}$ contributions taken into account in the PS scheme, as argued and
demonstrated in Ref. \cite{mnp}. It turns out that the effect due to  $\Obig{\alpha^2 L}$ corrections, which
can be inferred from the cross section values according to the following formula
\begin{equation*}
\delta_{\alpha^2 L}=100\times\dfrac{\sigma_{\textrm{exp}} - \sigma_{\alpha}^{\textrm{NLO}} - 
\sigma_{\textrm{exp}}^{\textrm{PS}}+\sigma_{\alpha}^{\textrm{PS}}}{\sigma} \ ,
\end{equation*}
does not exceed the 0.05\% level. 

As a whole, these results demonstrate that both next-to-leading $\Obig{\alpha}$ and 
multiple photon corrections are unavoidable for $0.1\%$ theoretical precision. 

\subsection{Differential distributions}
\begin{figure}[!htb]
\begin{center}
\includegraphics[width=285pt]{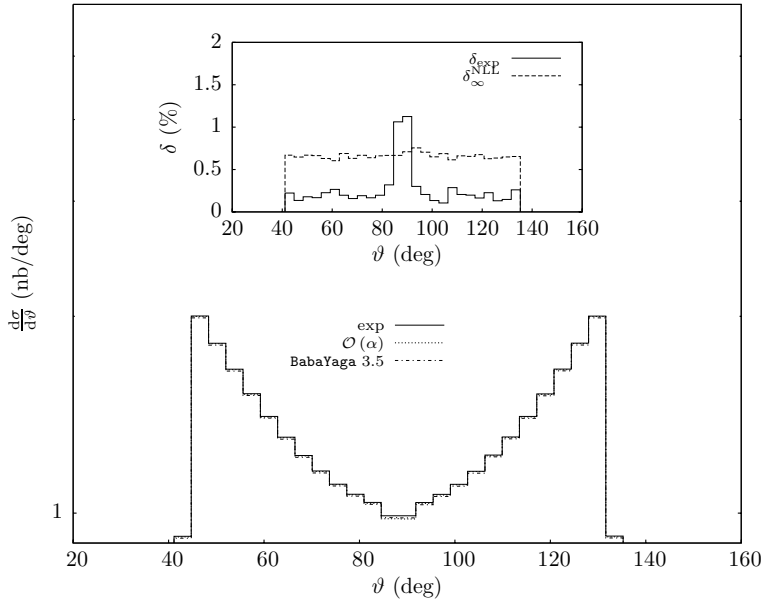}
\end{center}
\caption{Angular distribution of the most energetic photon according to the PS matched with $\Obig{\alpha}$ corrections (Eq. (\ref{eq:6}), solid line), the exact $\Obig{\alpha}$ 
calculation (Eq. (\ref{eq:3}), dashed line) and the pure all-order PS as in BabaYaga v3.5 
(dash-dotted line). lnset: relative effect (in per cent) of multiple photon corrections (solid line) 
and of non-log contributions of the matched PS algorithm (dashed line).}
\label{fig:lead_th}
\end{figure}

\begin{figure}[!htb]
\begin{center}
\includegraphics[width=285pt]{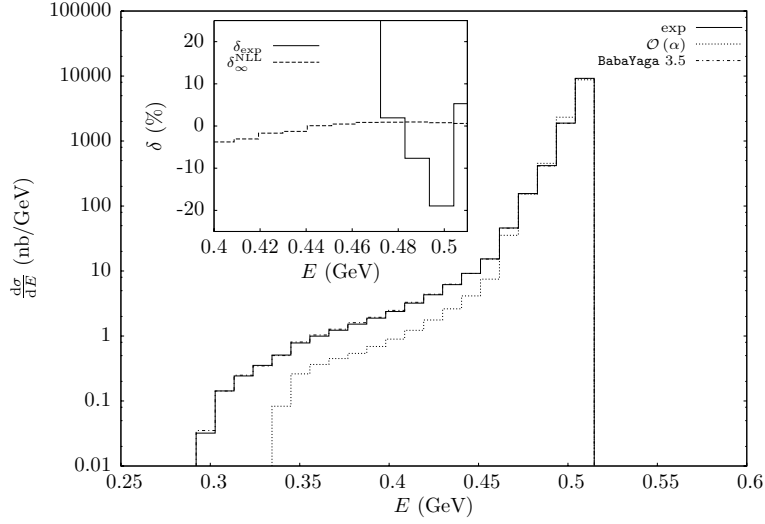}
\end{center}
\caption{The same as Fig. 3 for the energy distribution of the most energetic photon.}
\label{fig:lead_en}
\end{figure}

\begin{figure}[!htb]
\begin{center}
\includegraphics[width=285pt]{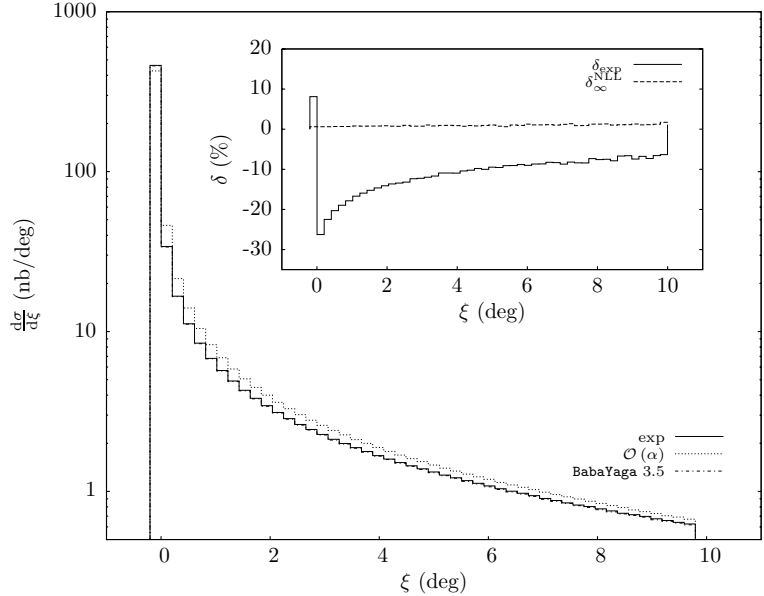}
\end{center}
\caption{The same as Fig. 3 for the acollinearity distribution of the two most energetic photons.} 
\label{fig:acoll}
\end{figure}

In Fig. \ref{fig:lead_th} and Fig. \ref{fig:lead_en} we show the angular and energy distribution 
of the most energetic photon,  while in Fig. \ref{fig:acoll} the acollinearity distribution of 
the two most energetic photons is represented. The above distributions, which have been 
simulated by using the latest version of the generator BabaYaga, correspond to the experimental
set up of Eq. (\ref{eq:KLOE}) and refer to exact  $\Obig{\alpha}$ corrections matched with 
the PS algorithm as in Eq. (\ref{eq:6}) (solid line), to the exact NLO calculation as in Eq.~(\ref{eq:3})  
 (dashed line) and to all-order pure PS predictions of BabaYaga v3.5~\cite{Ca} 
 (dash-dotted line). In the inset of each plot, the relative effect due to multiple photon contributions
 ($\delta_{\textrm{exp}}$) and non-logarithmic terms entering the improved PS algorithm
($\delta_{\infty}^{\textrm{NLL}}$) is also shown, according to the definitions given in 
Eq.~(\ref{eq:deltas}). 
 
For the angular distribution of the most energetic photon, the contribution of higher-order 
corrections beyond $\Obig{\alpha}$ amounts to about 1\% in the central region and is at 
a few per mille level at the edges of the distribution. More pronounced effects due to exponentiation
are present for the the energy distribution of the most energetic photon. In the statistically 
dominant region around 0.5~GeV, higher-order corrections reduce the $\Obig{\alpha}$ 
distribution of about 20\%, while they give rise to a significant hard tail in the proximity 
of the energy threshold of 0.3$\sqrt{s}$, as a consequence of the higher photon multiplicity 
of the resummed calculation with respect to the fixed  $\Obig{\alpha}$ prediction. Concerning
the acollinearity distribution, the contribution of higher-order corrections is positive and of about
10\% in correspondence of quasi back-to-back photon events, whereas it is negative and decreasing
from $\sim -30$\% to $\sim -10$\% for increasing acollinearity values.

As far as the contributions of non-logaritmic effects, dominated by next-to-leading  $\Obig{\alpha}$
corrections, are concerned, they contribute at the level of some per mille for the angular and acollinearity distribution, while they lye in the some per cent range for the energy distribution.

Therefore, as for the cross section, the interplay between $\Obig{\alpha}$ corrections and
exponentiation is crucial for precise predictions at the level of differential cross sections.

\section{Conclusions}
We have presented a high-precision QED calculation of the $e^+ e^- \rightarrow \gamma\gamma$ 
process, of interest for luminosity monitoring at flavour factories. The calculation, which includes 
all the relevant radiative corrections at the 0.1\% precision level, is implemented in an improved
version of the event generator BabaYaga\footnote{The code is available at
\texttt{http://www.pv.infn.it/\~\,\!hepcomplex/babayaga.html}.}, in order to contribute to a reduction of the
uncertainty in luminosity measurements at $e^+ e^-$ colliders 
of moderately high energies. The accuracy of the
approach has been demonstrated through a careful analysis of the various sources of 
photonic corrections to the experimentally relevant observables. The per mille precision reached 
for the two photon production process can, indeed, take advantage of the absence in such a process
of the vacuum polarization uncertainty, present in both $e^+ e^- \to \mu^+ \mu^-$ and 
 $e^+ e^- \to e^+ e^-$, and can also rely on previous estimates of sub-leading two-loop corrections 
 to Bhabha scattering,  that have been shown to typically contribute at the level of
 a few 0.01\% in the energy range explored at flavour factories (see {\it e.g.} 
 Ref. \cite{B1} and references therein).
 
 Possible perspectives concern the application of the approach, here presented for 
 $e^+ e^- \rightarrow \gamma\gamma$ and in Ref. \cite{B1} for the Bhabha process, to
obtain precise predictions for $e^+ e^- \to \mu^+ \mu^-$ and $e^+ e^- \to \mu^+ \mu^-\gamma$, 
 both of interest for physics studies at flavour factories, as well as the inclusion of the exact 
  $\Obig{\alpha}$ weak corrections to the two photon production process. 

\vskip 12pt\noindent
{\bf Acknowledgments}\\
We are grateful to H. Czy\.z, A. Denig,  S. Eidelman, S. Muller, F. Nguyen and G.~Venanzoni
for useful discussions and interest in our work. C.M. Carloni Calame is supported
by an INFN post-doc fellowship and is grateful to the University of Southampton for hospitality. 
This work is partially supported by the INTAS project Nr 05-1000008-8328 ``Higher-order effects in
  $e^+e^-$ annihilation and muon anomalous magnetic moment''.

\end{document}